\documentclass[sigconf]{acmart}

\usepackage{algorithm}
\usepackage{algpseudocode}
\usepackage{subfig}
\usepackage{enumitem}
\usepackage{graphicx}
\usepackage{multicol}
\usepackage{amssymb}
\usepackage{hhline}
\usepackage[labelfont=bf]{caption}
\usepackage{subfig}
\usepackage{float}
\usepackage{color}
\usepackage{graphicx}
\usepackage{multirow}
\usepackage{tabularx}
\usepackage{array,graphicx}
\usepackage{booktabs}
\usepackage{pifont}
\usepackage[utf8]{inputenc}
\usepackage[english]{babel}
\usepackage{amsmath}
\theoremstyle{definition}
\newtheorem{defn}{Definition}[section]



\copyrightyear{2018}
\acmYear{2018}
\setcopyright{acmcopyright}
\acmConference[HT '18]{29th ACM Conference on Hypertext and Social
	Media}{July 9--12, 2018}{Baltimore, MD, USA}
\acmPrice{15.00}
\acmDOI{10.1145/3209542.3209552}
\acmISBN{978-1-4503-5427-1/18/07}

\begin{document}
\title{Securing Social Media User Data - An Adversarial Approach}

\author{Ghazaleh Beigi}
\affiliation{%
	\institution{Arizona State University}
}
\email{gbeigi@asu.edu}

\author{Kai Shu}
\affiliation{%
	\institution{Arizona State University}
}
\email{kai.shu@asu.edu}

\author{Yanchao Zhang}
\affiliation{%
	\institution{Arizona State University}
}
\email{yczhang@asu.edu}

\author{Huan Liu}
\affiliation{%
	\institution{Arizona State University}
}
\email{huan.liu@asu.edu}

\renewcommand{\shortauthors}{G. Beigi et al.}

\begin{abstract}
Social media users generate tremendous amounts of data. To better serve users, it is required to share the user-related data among researchers, advertisers and application developers. Publishing such data would raise more concerns on user privacy. To encourage data sharing and mitigate user privacy concerns, a number of anonymization and de-anonymization algorithms have been developed to help protect privacy of social media users. In this work, we propose a new adversarial attack specialized for social media data. We further provide a principled way to assess effectiveness of anonymizing different aspects of social media data. Our work sheds light on new privacy risks in social media data due to innate heterogeneity of user-generated data which require striking balance between sharing user data and protecting user privacy.
\end{abstract}

%
%

\maketitle

\section{Introduction}
Explosive growth of social media in the last decade has drastically changed the web and billions of people all around the globe can freely conduct numerous activities such
as creating online profiles, interacting with other people, sharing posts, and various personal information in a rich \textit{heterogeneous} environment~\cite{beigi2018similar,gbeigi}. The resulted user-generated social media data consists of different aspects such as links, posts and profile information. This data provides opportunities for researchers and business partners to study and understand individuals at unprecedented scales~\cite{backstrom2007wherefore,hajibagheri2017extracting,beigi2016exploiting}.

However, publishing social media network data risks exposing people's privacy as the data is rich in content and relationship and contains individuals' sensitive and private information, resulting in privacy leakage~\cite{qian2016anonymizing,narayanan2009anonymizing,ji2016general}. For example, users' sensitive information such as vacation plans and medical conditions can be easily inferred from their posts. Publishing complete and intact social media data could even result in inferring sensitive information the users do not explicitly disclose such as age and location~\cite{beretta2015interactive}.

Privacy issues of users mandate social media data publishers to protect users' privacy by anonymizing the data. One straightforward anonymization technique is to remove ``Personally Identifiable Information'' (a.k.a. PII) such as names, user ID, age and location information and keep the social graph structure as is. This solution has been shown to be far from sufficient to protects people's privacy~\cite{narayanan2009anonymizing,backstrom2007wherefore}. An example of this insufficient approach is the anonymized dataset published for the Netflix prize challenge. Later, the work of~\cite{narayanan2008robust} showed that the structure of the data carried enough information for a potential breach of privacy to re-identify anonymized users. Consequently, various protection techniques have been proposed for anonymizing each aspect of the heterogeneous social media data. For example, some works perform anonymization on graph data structure~\cite{dwork2008differential,liu2008towards}, and others anonymize users' location information~\cite{puttaswamy2014preserving}. 
In general, the ultimate goal of an anonymization approach is to preserve social network user privacy while ensuring the utility of published data. 

Existing anonymization techniques often make a specific assumption regarding the way social media data is anonymized. In particular, these works assume that it's enough to anonymize each aspect of heterogeneous social media data (e.g., structure, textual, and location information) independently. At the first glance, this assumption makes sense as anonymization takes time and effort. Moreover, users privacy is protected while the data utility is preserved at the highest possible level. For example, lets consider the simplest case study in which published data includes only two aspects such as (i) structural (e.g., friendship, follower/followee links) and (ii) textual (e.g., posts) information.
We will then have options as shown in Table \ref{Tab:Hypothesis} to anonymize the data: no anonymization for either aspect, anonymization for one aspect, and anonymization for both. To ensure anonymization efficiency, as each aspect can be of different data types, a common practice is to anonymize each aspect independently. With two aspects as shown in Table \ref{Tab:Hypothesis}, case 4 is the backbone of the anonymization techniques for publishing data which is clearly the strongest protection of privacy. 

Privacy advocates have argued that sensitive information could be still leaked from the dataset anonymized considering each of these cases, but we lack conclusive evidence. It is unclear how the latent relation between different aspects of the data could be captured, whether the sensitive information with the scale of millions of users could be still leaked and what the success rate of such an attack could be. In particular, in this research, we are interested to study these issues by answering the following research questions:
\begin{itemize}[leftmargin=*]
	\item \textbf{(RQ1)}: Is the data private if just one of its two aspects is anonymized?
	\item \textbf{(RQ2)}: Is case 4, the strongest among four cases, sufficient for anonymizing social media data? 
\end{itemize} 
\begin{table}[t]\vspace{-0.20cm}
	\centering
	\small
	\caption{\textbf{Four different cases for social media data anonymization. Each check mark corresponds to the aspect of data being anonymized.}}
	\begin{tabular}{|l|l|l|l|l|}
		\hline
		& Case 1 & Case 2 & Case 3& Case 4\\ \hhline{=====}
		Structural Anonymization & \ding{55} & \ding{55} & \checkmark & \checkmark\\ 
		Textual Anonymization &  \ding{55} & \checkmark & \ding{55} & \checkmark\\ \hline
	\end{tabular}\label{Tab:Hypothesis}
	\vspace{-0.2cm}
\end{table}
Following the work of~\cite{narayanan2009anonymizing}, we seek to answer these questions by taking an adversary approach to assay the privacy level of anonymized social media data. However, existing de-anonymization attacks require a list of target users. A target user is an individual $v$ with the known identity in social media network $\mathcal{T}$ which will be mapped to a user in the given anonymized dataset. These techniques also require background knowledge $\mathcal{B}_v$ for each targeted user $v$ before initiating the attack. These methods require time and effort to find a proper set of target users and gather their knowledge which may not be realistic in practice. To address these challenges, we first introduce a new generation of adversarial attacks specialized for social media data which does not require collecting information before initiating the attack. Furthermore, to assess different ways of the social media dataset anonymization and answer the aforementioned questions, we propose a novel Adversarial Technique for Heterogeneous Data, namely, \textsc{Athd} which utilizes the latent relationship between different aspects of data. This new approach particularly well suits for social media data in which it is concerned with assessing the strengths of anonymizing different aspects of data. Our contributions could be summarized as follows:
\begin{itemize}[leftmargin=*]
	\item We introduce a new generation of adversarial attacks applicable to social media network data.
	\item We propose a novel de-anonymization technique \textsc{Athd} to assess the privacy level of anonymized heterogeneous social media data.
	\item We implement and evaluate \textsc{Athd} on two real world datasets to study the strengths of anonymization techniques in context of heterogeneous social media data. Our results demonstrates hidden relations between different aspects of the heterogeneous data make data anonymization techniques inefficient.
\end{itemize}


\section{Background}\label{preliminaries}
In this section, we review the technical preliminaries of protecting user privacy in social media data, i.e. data anonymization, which is required for the rest of this discussion.
Without loss of generality, in this paper, we assume that the published social media data consists of two aspects, namely, structure and textual information. 
More formally, we model the social network data as $\mathcal{D} = \left(\mathcal{V}, \mathcal{E}, \mathcal{P}\right)$ where $\mathcal{V} = \{\mathit{i} | \mathit{i} \text{ is a node}\}$ is the set of nodes or users, $\mathcal{E} = \{ \mathit{e}_{\mathit{i,j}} | \mathit{i, j} \in \mathcal{V}  \wedge$ there is a link from user $\mathit{i} \text{ to user } \mathit{j}\}$ is the set of links between any two nodes in $\mathcal{V}$ (e.g., friend and follower/followee relations), and $\mathcal{P}=\{\mathcal{P}_\mathit{i} | \mathit{i} \in \mathcal{V}\}$ is the set of all posts (textual information) associated with users in $\mathcal{V}$. $\mathcal{P}_\mathit{i} = \{ \mathit{p^{i}_1}, \mathit{p^{i}_{2}}, ..., \mathit{p^{i}_{m_i}} \}$ denotes posts by user $\mathit{i}$ where $m_i$ is the number of posts for user $\textit{i}$. Note that links in social networks could be either directed (e.g., follower/followee relation in Twitter) or indirected (e.g., friend relation in Facebook). We focus on directed graphs, although it is straightforward to apply the settings on undirected graphs as well. In order to preserve users' privacy, data publisher should anonymize the social media data $\mathcal{D}$ using privacy preservation techniques. Next, we will discuss techniques deployed to secure structural and textual information.
\subsection{Structural Information Anonymization}
To anonymize structural information, we first remove users' personally identifiable information (PII) such as user's name and ID. Techniques such as $k$-degree anonymity~\cite{liu2008towards}, sparsification, perturbation and switching~\cite{ji2015secgraph} are used for adding or removing nodes and links. The aim of $k$-anonymity methods is to anonymize each node so that it is indistinguishable from at least $k-1$ other nodes~\cite{sweeney2002k}. Liu~\textit{et al.} proposed to achieve $k$-degree anonymization~\cite{liu2008towards} through edge addition/deletion strategies~\cite{liu2008towards}. Sparsification technique randomly removes a set of $p|\mathcal{E}|$ edges ($p$ is the anonymiztion coefficient) while switching methods switches $\frac{p|\mathcal{E}|}{2}$ pairs of edges. Perturbation approach first removes a set of $p|\mathcal{E}|$ edges and then add same amount of edges randomly~\cite{ji2015secgraph}.
\subsection{Textual Information Anonymization}\label{text-anonymiziation}
In this work, we anonymize the textual information using $\epsilon$-differential privacy~\cite{dwork2008differential} by first converting each user's post into a numerical vector using tokenizing and calculating Term Frequency Inverse Document Frequency (TF-IDF) scores and then adding Laplacian noise to the text vector. Details are discussed next.
\subsubsection{Text Processing}\label{text_processing}
To anonymize user $\mathit{i}$'s posts, we first remove user's PII such as user ID (including mentioning and retweeting), name and link information from her texts. Then, we follow a standard process to convert each of user's posts to a numerical vector. To do so, we first consider posts by all users in the dataset and perform some pre-processing including stop word removal. The unigram model is then deployed to construct the word feature space $\mathcal{W}$. Finally, we use Term Frequency Inverse Document Frequency (TF-IDF) as a feature weight to derive the vector $\textbf{x}^{i}_{l}$ for each post $\mathit{p^{i}_l}$ of user $\mathit{i}$. TF-IDF score for each word $t$ is calculated as:
\vspace{-8pt}
\begin{equation}
\mathbf{x}^{i}_{l}\left(t\right) = f^{i}_l(t)*\log \frac{M}{n_t}
\end{equation}
where, $f^{i}_{l}(t)$ is the number of times word $t$ appeared in the post $\mathit{p^{i}_{l}}$, $M$ is the total number of posts in the data and $n_t$ is the number of posts that the word $t$ was used in them. We can represent $\mathit{p^{i}_{l}}$ with the corresponding vector $\mathbf{x}^{i}_{l}$. All users' posts can be then denoted by the post-word matrix $\mathbf{X} \in \mathbb{R}^{M \times \lvert\mathcal{W}\rvert}$ where $\lvert\mathcal{W}\rvert$ denotes the size of the word space. Relations between users and posts can be also represented via a user-post matrix $\mathbf{W} \in \mathbb{R}^{N \times M}$ where $N$ is the number of users and $\mathbf{W}_{ij}=1$ if post $j$ was posted by user $i$ and $\mathbf{W}_{ij}=0$ otherwise. Next, we will discuss how we leverage differential privacy technique to anonymize the textual information.
\subsubsection{Differential Privacy}
We use differential privacy technique proposed in~\cite{dwork2008differential} to anonymize the textual information. Differential privacy aims at maximizing privacy of users when a statistical query is submitted over a database and an answer is retrieved. Formally, $\epsilon$-differential privacy is defined as follows:
\begin{defn}{\textbf{ ($\epsilon$-Differential Privacy~\cite{dwork2008differential})}}.Given an input dataset $\mathcal{H}$, a query $f(\mathcal{H})$ and a desirable output range, a mechanism $K(.)$ with an output range $\mathcal{R}$ satisfies $\epsilon$-differential privacy iff,
	\begin{equation}
	\frac{Pr[K(f(\mathcal{H}_1)=R \in \mathcal{R})]}{Pr[K(f(\mathcal{H}_2)=R \in \mathcal{R})]}\leq e^\epsilon
	\end{equation}
for any datasets $\mathcal{H}_1, \mathcal{H}_2$ that differ in only one row, where $\epsilon$ is the privacy budget.
\end{defn}

Larger values of $\epsilon$ result in a larger privacy loss as the changes of the database can be inferred more easily, and, smaller values of $\epsilon$ lead to smaller privacy loss and a higher tolerance of database to privacy breach. Note that the differential privacy is just a condition on a mechanism which releases the dataset. The mechanism which achieves $\epsilon$-differential privacy is called sanitization. Laplacian mechanism is one popular sanitization technique which gives differential privacy for real valued queries by adding a Laplacian noise~\cite{dwork2008differential}. Assume that $f(\mathcal{H})$ is the real value response to a certain query $f$. Then, a random noise $Y(\mathcal{H})$ is generated from Laplacian distribution and added to $f(\mathcal{H})$ as:
\begin{equation}\label{noise2}
K\left(f\left(\mathcal{H}\right)\right) = f\left(\mathcal{H}\right) + Y\left(\mathcal{H}\right)
\end{equation}

The Laplacian distribution has zero mean and a scale parameter $\Delta(f)/\epsilon$  where $\Delta(f)$ is the sensitivity of $f$ and defined as the maximum variation of the query function between datasets differing in at most one record:
\begin{equation}
\Delta(f) = \max \lVert f(\mathcal{H}_1) - f(\mathcal{H}_2)\rVert _1
\end{equation}
The density function of the Laplacian noise will be computed as:
\begin{equation}
p(x) = \frac{\epsilon}{2\Delta(f)} e ^{-\frac{\lvert x\rvert \epsilon}{\Delta(f)}}
\end{equation}

Note that higher sensitivity $\Delta(f)$ of the query function $f$ with fixed $\epsilon$, implies more Laplacian noise added to $f(\mathcal{H})$. 
\subsubsection{Anonymizing Textual Information with Differential Privacy}
In order to anonymize the post-word matrix $\mathbf{X}$ in a way that $\epsilon$-differential privacy is preserved, we need to apply the discussed mechanism $K(.)$ on the original matrix $\mathbf{X}$ and transform it into a new one $X' = K(X)$. Instead of transforming the entire matrix $\mathbf{X}$ at once, we can transform each individual row of the matrix by adding a Laplacian noise to $\mathbf{X}_i$ to create a new row $\mathbf{X'}_i$. Considering the identity query function $f_I(\cdot)$ where $f(D)=D$, the sensitivity of $f_I(\cdot)$ can be defined as follows:
\begin{equation}
\Delta(f_I) = \max\lVert\mathbf{X}_i - \mathbf{X}_j\rVert _1
\end{equation}
where $\mathbf{X}_i$ and $\mathbf{X}_j$ are any two random row vectors from $\mathbf{X}$. Following the equation~\ref{noise2}, a Laplacian noise will be added to each vector $\mathbf{X}_i$:
\begin{equation}\label{noise}
K(f_I(\mathbf{X}_i)) = \mathbf{X}_i + \left[Y_{i1},...Y_{i|\mathcal{W}|}\right], i = 1,...,n
\end{equation}
Similarly, $Y_{ij}$'s are drawn i.i.d. from Laplacian distribution with zero mean and $\Delta(f_I)/\epsilon$ scale parameter. After anonymizing the textual information, the anonymized post-word $\mathbf{X}$ and user-post $\mathbf{W}$ matrices will be published. The information regrading the word feature space $\mathcal{W}$ will be released by the data publisher as well. 


\section{Social Media Adversarial Attack}\label{Threat}
\begin{figure}[ht]
	\vspace{-20pt}
	\centering 
	\subfloat[\bf{Traditional de-anonymization.}]{\includegraphics[width=0.4\textwidth]{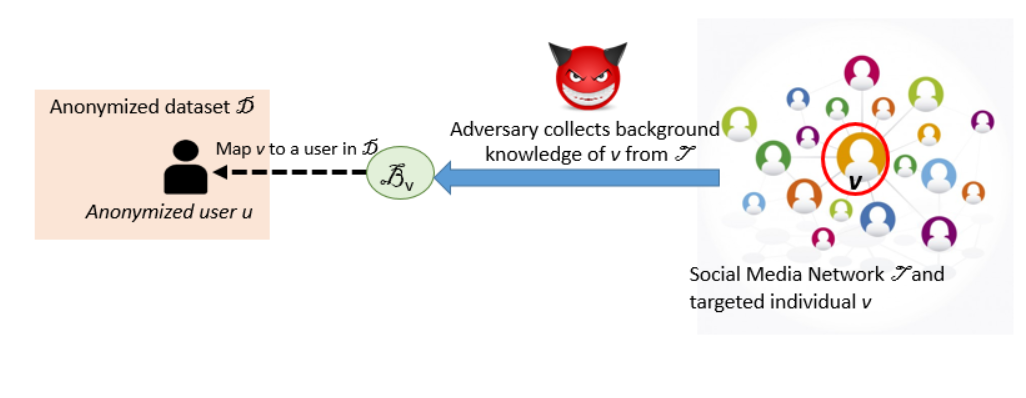}}\\
	\vspace{-10pt}
	\subfloat[\bf{Proposed social media adversarial attack}]{\includegraphics[width=0.4\textwidth]{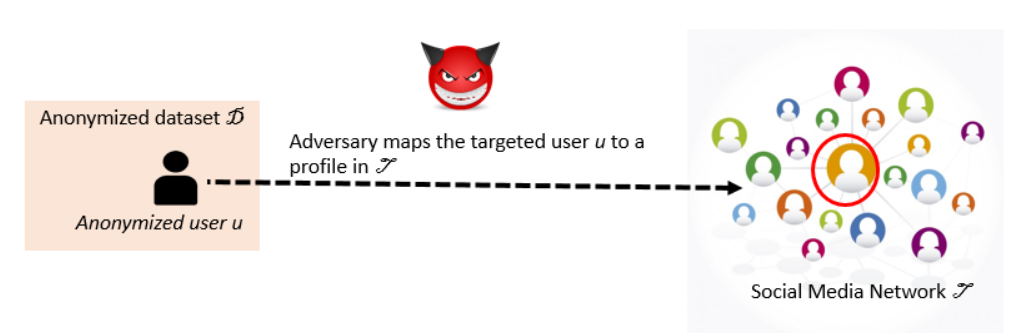}}
	\caption{\textbf{Traditional de-anonymization vs. proposed social media adversarial attack.}}\label{attack}\vspace{-0.4cm}
\end{figure}

De-anonymization techniques have been proposed in the literature as a counterpart to data anonymization research direction~\cite{yartseva2013performance,pedarsani2013bayesian,ji2016general,fu2015effective,qian2016anonymizing}. De-anonymization works further help improve anonymization techniques and reduce privacy breach by probing the potential drawbacks of anonymization techniques. Figure~\ref{attack}(a) depicts how these de-anonymization approaches work. These works assume that the adversary has been given a list of target users to de-anonymize requiring adversarial to collect background knowledge about target users before initiating the attack~\cite{abawajy2016privacy}. 

Narayanan et.al.~\cite{narayanan2009anonymizing} discuss different ways of collecting background knowledge such as crawling data via social media networks API. Since these methods require time and effort to gather knowledge, it may not be realistic in practice for two reasons: (1) the number of target users can be very large, thinking about the number of users in Twitter; and (2) most of the online social media APIs have rate limits on the number of request a user can make through their APIs in a specific time window. Also, these APIs can only provide a random small portion of available data for each search query. 
This makes it infeasible to collect the background information for a significant number of users in $\mathcal{T}$ in order to find the one-to-one mapping between users in $\mathcal{D}$ and $\mathcal{T}$. Therefore, the above target-user-based approach cannot be applied to social media users when no list of target user is given. 

To address these shortcomings, we introduce a new generation of adversarial attacks (Figure~\ref{attack}(b)) specialized for social media network data. This approach does not require the attacker to gather background knowledge $\mathcal{B}$ before starting the attack. In fact, users registered in social media which are available via online APIs are the adversaries' only source of information. The adversary can send queries to these APIs, anytime during the adversarial process. It is formally defined bellow:\vspace{-0.15cm}
\begin{defn}{\textbf{(Social Media Adversarial Attack)}}. Given an anonymized social media network dataset $\mathcal{D}$, the aim of adversarial attack is to find a one-to-one mapping between each user $\textit{u}$ in $\mathcal{D}$ and a real identity in targeted online social media network $\mathcal{T}$.
\end{defn}\vspace{-0.1cm}
Next, we will accordingly discuss the details of our proposed de-anonymization approach, \textsc{Athd}, which does not require collecting target users and their background information and is proposed to further evaluate heterogeneous social media anonymization.


\section{Adversarial Technique for Heterogeneous Data}\label{de-anony}
Our proposed de-anonymization technique, adversarial technique for heterogeneous data (\textsc{Athd}), uses different aspects of data, i.e., graph structure and users' textual information to identify the real identity of users in the anonymized dataset $\mathcal{D} =  (\mathcal{V}, \mathcal{E}, \mathbf{X}, \mathbf{W}, \mathcal{W})$. We posit that this attack could be applied on various aspects of data and is not limited to only two data aspects or structural and textual information.
%
	
 The main idea behind de-anonymization is to find the most similar user in social media $\mathcal{T}$ to the user $u$ in the anonymized dataset. Here, we follow the same approach as the existing works, meanwhile our goal is to design a new framework which exploits the hidden relations between different aspects of the data, to eventually map the users to their real profile in $\mathcal{T}$.

Our proposed de-anonymization consists of three main steps. 
Given the anonymized dataset $\mathcal{D}$, we first extract the most revealing information for $\textit{u}$. Second, we search those information in search engine of the targeted social media $\mathcal{T}$. This search returns a list of people whose posts include the inquired query. We save all the returned candidates as a candidate set. Third, we identify the profile from the candidate set most similar to the user $\textit{u}$ . The details of each of three steps are discussed next.
\subsection{Step 1: Extracting the Most Revealing Information}
The first step includes extracting the most revealing information for user $\textit{u}$ via social media API. In this work, we rather use textual information since it is not straightforward to look up information related to links. We are thus interested in extracting the most revealing textual information of user $\textit{u}$. We assign a score $s_l$ to each post $l$ of $u$, $\{l \in \{1,...,M\}| \mathbf{W}_{ul}=1\}$ to measure how unique each post $l$ is. Each post $l$ has been vectorized using tf-idf approach and is represented in $l$-th row of the post-word matrix $\mathbf{X}$. Given the vector representation $\mathbf{X}_l$ of post $l$, the score $s_\textit{l}$ is calculated as,
\begin{equation}\label{score}
s_\mathit{l} = \frac{\sum_{t=1}^{\mathcal{W}} \mathbf{X}_{l}(t)}{\lvert\mathcal{W}\rvert}
\end{equation}
The higher this score is, the more unique and thus the more revealing post $\textit{l}$ would be. Based on this, we rank user $\textit{u}$'s posts and select the top-$k$ posts as the most revealing information. 
\subsection{Step 2: Finding a Set of Candidates}\label{candidate}
The goal of this step is to find a set of candidates for each user $\textit{u}$, given the top-$k$ most revealing posts. To do so, for each nominated post $l$ from step 1, we select set of words $\mathcal{S}$ whose tf-idf scores are greater than the average of the tf-idf scores for the words in the post $l$, $\mathcal{S}^l = \{t |\mathbf{X}_{l}(t)>s_\mathit{l}\}$. This approach helps to not to select useless words which have non-zero tf-idf values only due to data distortion during the anonymization process. Therefore, the words with higher chances of being posted in a real text are selected. This step results in a set of queries $\mathcal{Q}_\textit{u} = \{q_u^{(1)}, q_u^{(2)}, ..., q_u^{(k)}\}$. We construct the query $q_u^{(i)}$ from set $\mathcal{S}^i$, $i \in \{1,...,k\}$ as $q_u^{(i)} = \{word \in \mathcal{S}^i\}$. Each of $ q_u^{(i)} \in \mathcal{Q}_\textit{u}$ is queried through the $\mathcal{T}$'s search engine. Result includes a set of users who have published posts including keywords in $ q_u^{(i)}$.

Integrating results from all queries in $\mathcal{Q}_\textit{u}$, we have a set of candidate users for user $\textit{u}$ which is denoted by $\mathcal{C} = \{c_1, c_2,...,c_{|\mathcal{C}|}\}$. Combining steps 1 and 2, we first find posts which are the most revealing for user $\textit{u}$ and then for each selected post we select the words that are more likely to be used by the same user. The result will be a set of candidates for $u$. 
\subsection{Step 3: Matching-Up Candidates to Target}
In the last step, we find the most similar candidate to user $\mathit{u}$. We shall define a metric which measures the similarity between each user $\textit{u}$ and $i$th candidate $c_i \in \mathcal{C}$. Previous works~\cite{narayanan2009anonymizing,ji2016graph,ji2016general,qian2016anonymizing,nilizadeh2014community} have solely leveraged the structural properties to find the similarity between a target user $v$ and users in an anonymized dataset. However, given the properly anonymized network, the attacker is not be able to accurately find the similarity between users by just incorporating structural properties. We use other aspects of the data (even if they are anonymized) along with structural properties to reveal interesting information that could be leveraged for inferring the similarity. Location, textual and profile information are good examples of such social media data aspects. We consider textual information as the second aspect of the data. We stress that out proposed approach is not limited to textual and structural information and could be generalized to any data type. 
We also assume that the adversary is not aware of details of deployed anonymization techniques. Next, we define two sets of features to calculate the similarity between $\textit{u}$ and her $i$-th candidate $c_i$. 
\vspace{-0.1cm}
\subsubsection{Structural Features.} \label{struct_similarity}
It has been also shown that users can be uniquely identified using their neighbors degree distributions~\cite{hay2007anonymizing}. Following previous works~\cite{sharad2016true}, we thus leverage degree distribitons of $u$'s neighbors $\mathcal{N}(u)$ (i.e. all followers and followees of $u$) in order to represent her structural features. 
Note that properties such as betweenness, closeness, and eigenvector centrality cannot be considered as $u$'s structural features since it requires having access to the complete network of users in $\mathcal{T}$ which is not feasible in practice. We quantify degree distributions by categorizing them into $b$ bins with size of $\delta$ in a way that each bin contains the number of neighbors that have the degree in assigned range of that bin. For directed graphs, neighbors of each user can be divided into two groups of follower and followee and final feature set is computed by concatenating result of each group.
\subsubsection{Textual Features.}\label{textual_similarity}
Remind that for user $\textit{u}$, the attacker is given a set of $m_u$ textual vectors as well as word space features $\mathcal{W}$. For each candidate $c_i$, we collect a set of $\theta$ recent posts by sending requests to the $\mathcal{T}$ API. The collected posts are then concatenated in one unified document. Next, $c_i$'s PII will be removed from the document and the corresponding text vector is then created given the word space $\mathcal{W}$ following the similar approach of the Section~\ref{text_processing}. Textual features for users $\textit{u}$ and $c_i$ are thus represented by a set $m_u=\{t_1, t_2, ...t_{m_u}\}$ and a textual vector $t_{c_i}$, respectively.
\subsubsection{Calculating Users Similarity.} 
Given two groups of structural and textual features, similarity between $\textit{u}$ and $c_i$ is computed as the linear combination of their textual and structural similarities,
\begin{equation}\label{simple1}
Sim\left(\mathit{u}, c_i\right) = 	\alpha Sim_{struct}\left(\mathit{u}, c_i\right) + 	(1-\alpha)Sim _{text}\left(\mathit{u}, c_i\right)
\end{equation}
where $\alpha$ controls the contribution of structural similarity. We further define $Sim_{struct}(\mathit{u}, c_i)$ as the cosine similarity between the two structural vectors computed as $
Sim_{struct}\left(\mathit{u}, c_i\right) = \cos \left(s_{u},s_{c_i}\right)$. Textual similarity between $\textit{u}$ and $c_i$ is also computed as the average of cosine similarity between $\forall t_j \in m_u$ and $t_{c_i}$,
\begin{equation}\label{simple2}
Sim _{text}(\mathit{u}, c_i) = \frac{\sum_{j = 1}^{|m_u|} \cos \left(t_{j},t_{c_i}\right)}{m_u}
\end{equation}
\subsubsection{Improving Similarity Measure.} Merely checking the structural and textual similarity between the two users' features may lead to biased and not accurate results. Moreover, the attacker needs a more powerful similarity metric which could reduce the effect of anonymization. To handle this issue, we follow a fundamental well-defined problem in the field of image processing~\cite{buades2005non}, \textit{image denoising}. Non-local mean filters are a traditional way to remove noise from image data~\cite{buades2005non}. This approach replaces a pixel's value with the weighted average of all other pixels around it. The amount of weighting for neighboring pixels is based on the degree of similarity between a small patch centered on that pixel and a small patch centered on the pixel being denoised~\cite{buades2005non}. Inspired by the idea behind non-local mean filters~\cite{buades2005non}, we use the feature values of other users similar to user $\textit{u}$ in order to reduce effect of anonymization. To apply this idea, we first need to find similar users to $\textit{u}$-- here is where the concept of \textit{homophily} comes in handy. Homophily is one of the most important social correlation theories which is also observed in social media and explains the tendency of individuals to associate and create relationship with similar ones~\cite{mcpherson2001birds,crandall2008feedback}. 

Following the similar idea to non-local means filtering, we leverage homophily and consider user $\textit{u}$'s neighbor set $\mathcal{N}(u)$ as set of similar users to her. Utilizing homophily also helps in capturing the hidden relations between different aspects of the data. We thus calculate the similarity between $\mathcal{N}(u)$ and neighbors set $\mathcal{N}(c_i)$ for candidate $c_i$. We first quantify the degree distributions for \textit{all} users in both neighbors set $\mathcal{N}(u)$ and $\mathcal{N}(c_i)$ as discussed earlier in Section \ref{struct_similarity}. The structural similarity of neighbors are then calculated based on the cosine similarity between $s_{\mathcal{N}(u)}$ and $s_{\mathcal{N}(c_i)}$.

Following the procedure introduced in Section \ref{textual_similarity}, we collect and concatenate $\theta$ recent posts for all neighbors in $\mathcal{N}(c_i)$. Textual similarity between $\mathcal{N}(u)$ and $\mathcal{N}(c_i)$ is then computed by taking average over the cosine similarities between textual vector of each user in $\mathcal{N}(u)$ and the textual vector of $t_{\mathcal{N}(c_i)}$. The total similarity between neighbors will be then calculated as follows,
\begin{align}\label{improved1}
Sim\left(\mathcal{N}(u), \mathcal{N}(c_i)\right) = \alpha Sim_{struct}\left(\mathcal{N}(u), \mathcal{N}(c_i)\right) \nonumber \\
~+ 	(1-\alpha)Sim _{text}\left(\mathcal{N}(u), \mathcal{N}(c_i)\right)
\end{align}

This metric quantifies the fitness of $\mathcal{N}(u)$ and $\mathcal{N}(c_i)$ as the similarity scores of their structural and textual properties. It reduces the effect of data anonymization and also aligns well with the assumption that if $\textit{u}$ and $c_i$ correspond to the same identity, their neighbors $\mathcal{N}(u)$ and $\mathcal{N}(c_i)$ should also match~\cite{fu2015effective}. Finally, the total similarity between $\textit{u}$ and $c_i$ can be computed as the combination of their individual similarity and the fitness of their neighbors:
\begin{equation}\label{TotalSim}
Sim_{total}(\mathit{u}, c_i) = 	\beta Sim\left(\mathit{u}, c_i\right) + 	(1-\beta)Sim\left(\mathcal{N}(u), \mathcal{N}(c_i)\right) 
\end{equation}

 We empirically find that random selection of $c_i$'s neighbors with the size $\lambda$ works well in our problem and we are not required to collect all neighbors information from $\mathcal{T}$'s API. This will make the de-anonymization approach more efficient. Note that many noise removal approaches have been designed for specific kinds of noise (e.g., Guassian noise) which could be used to remove the noise from the data and particularly the vector of textual information. However, using certain noise removal approaches may not always have positive effects. In fact, it can lead to a wrong estimation of users' properties when the attacker does not have any prior knowledge of the deployed anonymized technique. 

\begin{algorithm}[t]
	\caption{\textbf{Adversarial Technique for Heterogeneous Data}}\label{alg:alg1}
	\begin{algorithmic}[1]
		\Require user $u$, Anonymized Data $\mathcal{D}$= \{$\mathcal{V}$, $\mathcal{E}$, $\mathbf{W}$, $\mathbf{X}$, $\mathcal{W}$\}, $k$, $\lambda$, $\theta$, $h$, $\alpha$, $\beta$
		\Ensure Top-$h$ mapped accounts in targeted social network $\mathcal{T}$
		\State Initialize the candidate set $\mathcal{C}=\phi$.
		\For {\text{each anonymized text vector of post $l$ for $u$}}
		\State Calculate score $s_l$ according to Eq.\ref{score}.
		\EndFor
		\State Select top-$k$ posts with the highest score $s_l$ as the most revealing information. 
		\For {\text For each text vector $l$ in top-$d$ posts}
		\State Select words with tf-idf scores $\mathbf{X}_{l}(t)>s_\mathit{l}$ to create search query $q_u^{(l)}$.
		\State Search query $q_u^{(l)}$ in $\mathcal{T}$ search engine and add results to $\mathcal{C}$.
		\EndFor
		\For {\text each candidate $c_i$ in $\mathcal{C}$}
		\State Calculate similarity between $\mathit{u}$ and $c_i$ according to Eq.\ref{TotalSim}.
		\EndFor
		\State \text{Return the top-$h$ candidates with maximum similarity}
	\end{algorithmic}
	
\end{algorithm}

The proposed \textsc{Athd} approach is shown in Algorithm \ref{alg:alg1}. The input to the algorithm is the anonymized dataset and the output is the top-$h$ mapped profile accounts in $T$. Lines 2--5, correspond to the first step of \textsc{Athd}. The set of candidate set (step 2) is then found through lines 6--9. The similarity between $\mathit{u}$ and each of the selected candidates is calculated in lines 10--12. Finally, top-$h$ candidates with the maximum similarity to $\mathit{u}$ will be returned. This re-identification procedure is then run over all users in the anonymized dataset. Note \textsc{Athd} is independent of deployed anonymization techniques either for the textual or structural information. In the next section we will discuss how our proposed de-anonymization could be generalized to the social media data with any type of components.

\subsection{Generalizability of \textsc{Athd}}\label{General}
Our framework can be generalized through abstraction to different social media data, assuming that our anonymized data consists of two different aspects, $\mathcal{A}_1$ and $\mathcal{A}_2$ and the attacker is willing to initiate an attack by mapping $\mathit{u}$ to a real profile in the targeted social media $\mathcal{T}$. As discussed before, the first step is to extract the most revealing information from $\mathcal{A}_1$ for user $u$ by using the same concept as tf-idf scores. 
 The second step includes selecting a set of candidate profiles for $u$ by searching for the extracted information from the previous step through $\mathcal{T}$'s search engine. Finally, the similarity between $u$ and her candidates are calculated using the combination of features of existing data components, $\mathcal{A}_1$ and $\mathcal{A}_2$. Features of the most similar users to $u$ (e.g., neighbors) are also incorporated as well to reduce the anonymization effect while capturing the hidden relation between different aspects of the data.

\section{Experiments}\label{expr}
In this section, we seek to answer the introduced research questions, but we first need to evaluate the efficiency of proposed adversarial technique \textsc{Athd}. We begin this section by introducing the dataset and anonymization techniques we used. Then, we compare the results of \textsc{Athd} against the state-of-the-art de-anonymization benchmarks to evaluate its effectiveness. Next, we use \textsc{Athd} to assess the anonymization power of each of the four cases to answer the research questions:
\begin{itemize}[leftmargin=*]
	\item \textbf{(RQ1)}: Is the data private if just one of its two aspects is anonymized?
	\item \textbf{(RQ2)}: Is case 4, the strongest among four cases, sufficient for anonymizing social media data? 
\end{itemize} 
\vspace{-10pt}
\subsection{Datasets}
We use two different datasets from two large social media websites, Twitter and Foursquare. Twitter is a prevalent and well-known microblogging social media allows millions of active users interacting with each other via short posts, called tweet. Foursquare is a location based social media in which users share their location with friends. Users can also leave tips about different places. We collect the Twitter dataset using Twitter API using the snowball sampling technique as follows. We begin with a random initial seed of users and for each user $u$ in the seed, we obtain a random subset of size 100 of her posted tweets as well as a subset of size 500 of her follower/followee information. We repeat the same process for each $u$'s followers/followees. This way we build our final dataset which consists of the users in the initial seed and their 2-hops connections. We follow the same procedure to collect the data from Foursquare API by considering a random initial seed of users. We collect each user friends as well as her tips on different locations. We build the final dataset by repeating this process for 2-hops connections. Note that in both datasets, we only keep the information of users who have posted at least one tweet or tip.

Next, we will apply various anonymization techniques on the obtained dataset-- this is described in the next section. Also, we utilize the Twitter's advanced search engine\footnote{https://twitter.com/search-advanced?lang=en} and Foursquare search \footnote{https://foursquare.com/explore?} during the de-anonymization process for Twitter and Foursquare data, respectively. It would be also worthwhile to add that we already have the ground truth for the re-identification, since the real profiles of the crawled users are known to us beforehand. Table \ref{tab:data_stat} summarizes the statistics of our datasets.
\begin{table}[t]

	\centering
	\small
	\caption{\textbf{Statistics of the crawled datasets.
		}}\label{tab:data_stat}
		\subfloat[\bf{Twitter}]{
			\begin{tabular}{l|l|l}
				\hline \hline
				\# of Users & \# of Edges & Avg. Clustering Coefficient\\ 
				6,789 & 244,480 & 0.219\\ \hline
				Density & \# of Tweets & \# of Unigrams \\		
				0.005 & 478,129 & 208,483\\ \hline \hline
			\end{tabular}}
			\quad
			\subfloat[\bf{Foursquare}]{
			\begin{tabular}{l|l|l}
				\hline \hline
				\# of Users & \# of Edges & Avg. Clustering Coefficient\\ 
				22,332 & 229,234 & 0.295\\ \hline
				Density & \# of Tips & \# of Unigrams \\		
				0.0005 & 124,744 & 103,264\\ \hline \hline
			\end{tabular}}\vspace{-0.7cm}
			\end{table}

\subsection{Anonymization Approaches}
We use different anonymization techniques to evaluate the introduced different anonymization cases in Table\ref{Tab:Hypothesis}. 
\begin{table*} [t]  \centering \small
		\subfloat[\bf{Twitter}]{
				\begin{tabular}{@{} clcc|cc|cc|cc @{}}
					& &  \multicolumn{2}{c}{\textbf{\textsc{Athd}-Improved}}& \multicolumn{2}{c}{\textbf{\textsc{Athd}-Simple}} & \multicolumn{2}{c}{\textbf{ADA}} & \multicolumn{2}{c}{\textbf{Narayanan et. al.}}\\[2ex]
					& & {\bf{{\footnotesize Naive}}} & {\bf{{\footnotesize Diff Privacy}}}& {\bf{{\footnotesize Naive}}} & {\bf{{\footnotesize Diff Privacy}}} & {\bf{{\footnotesize Naive}}} & {\bf{{\footnotesize Diff Privacy}}} & {\bf{{\footnotesize Naive}}} & {\bf{{\footnotesize Diff Privacy}}}\\
					\cmidrule{2-10}
					& \bf{Naive}  & 0.9435 (1) & 0.8020 (2) & 0.8200 (1) & 0.6951 (2)& 0.6729 (1)& 0.5513 (2)& 0.5073 (1)& 0.4100 (2)\\
					& \bf{Sparsification} & 0.8087 (3) & 0.6998 (4) & 0.7327 (3)& 0.6213 (4)& 0.6099(3) & 0.5114 (4)& 0.4316 (3)& 0.3437 (4)\\
					& \bf{$k$-deg(add)} & 0.7894 (3) & 0.6814 (4)& 0.6900 (3)& 0.6125 (4)& 0.5898 (3)& 0.4982 (4)& 0.3979 (3) & 0.3139 (4)\\
					& \bf{$k$-deg(add \& del)}  & 0.7580 (3) & 0.6533 (4)& 0.6891 (3)& 0.5821 (4)& 0.5800 (3)& 0.4727 (4) & 0.3815 (3) & 0.2997 (4)\\
					& \bf{Switching}  & 0.6911 (3) & 0.5812 (4)& 0.6013 (3) & 0.5186 (4)& 0.4971 (3) & 0.4014 (4) & 0.3520 (3) & 0.2618 (4)\\
					& \bf{Perturbation} & 0.6500 (3)& 0.5685 (4) & 0.5367 (3) & 0.4249 (4)& 0.4322 (3)& 0.3618 (4) & 0.2987 (3)& 0.2018 (4)\\
					\cmidrule[1pt]{2-10}
				\end{tabular}}
			\quad
			\subfloat[\bf{Foursquare}]{	\begin{tabular}{@{} clcc|cc|cc|cc @{}}
					& &  \multicolumn{2}{c}{\textbf{\textsc{Athd}-Improved}}& \multicolumn{2}{c}{\textbf{\textsc{Athd}-Simple}} & \multicolumn{2}{c}{\textbf{ADA}} & \multicolumn{2}{c}{\textbf{Narayanan et. al.}}\\[2ex]
					& & {\bf{{\footnotesize Naive}}} & {\bf{{\footnotesize Diff Privacy}}}& {\bf{{\footnotesize Naive}}} & {\bf{{\footnotesize Diff Privacy}}} & {\bf{{\footnotesize Naive}}} & {\bf{{\footnotesize Diff Privacy}}} & {\bf{{\footnotesize Naive}}} & {\bf{{\footnotesize Diff Privacy}}}\\
					\cmidrule{2-10}
					& \bf{Naive}  & 0.8004 (1) & 0.6799 (2) & 0.7107 (1) & 0.5989 (2)& 0.5699 (1)& 0.4821 (2)& 0.4400 (1)& 0.3754 (2)\\
					& \bf{Sparsification} & 0.7238 (3) & 0.6299 (4) & 0.6400 (3)& 0.5499 (4)& 0.5118(3) & 0.4532 (4)& 0.3968 (3)& 0.3028 (4)\\
					& \bf{$k$-deg(add)} & 0.6947 (3) & 0.5999 (4)& 0.6112 (3)& 0.5288 (4)& 0.5138 (3)& 0.4157 (4)& 0.3487 (3) & 0.2748 (4)\\
					& \bf{$k$-deg(add \& del)}  & 0.6612 (3) & 0.5739 (4)& 0.5918 (3)& 0.4989 (4)& 0.4867 (3)& 0.3947 (4) & 0.3025 (3) & 0.2639 (4)\\
					& \bf{Switching}  & 0.6134 (3) & 0.5431 (4)& 0.5517 (3) & 0.4614 (4)& 0.4300 (3) & 0.3521 (4) & 0.2987 (3) & 0.2120 (4)\\
					& \bf{Perturbation} & 0.5642 (3)& 0.4930 (4) & 0.4518 (3) & 0.3670 (4)& 0.3402 (3)& 0.2836 (4) & 0.2300 (3)& 0.1876 (4)\\
					\cmidrule[1pt]{2-10}
				\end{tabular}}

	\caption{\textbf{Comparison of the de-anonymization success rates for various anonymization techniques. Higher values imply higher privacy breach. Numbers in parentheses demonstrate the corresponding case number in Table \ref{Tab:Hypothesis}.}}\label{Res2}\vspace{-0.7cm}
\end{table*}
Following previous work~\cite{fu2015effective}, we choose different algorithms for \textit{structural information} anonymization as follows:
\begin{itemize}[leftmargin=*]
	\item \textbf{Naive Anonymization}. This approach only masks users' identifiers (PII), and does not change the graph structure. This is the simplest approach and thus we would
	expect the highest vulnerability and hence best de-anonymization result.
	\item \textbf{Sparsification}. This work randomly eliminates $p|\mathcal{E}|$ edges where $p$ is the anonymiztion coefficient.
	\item \textbf{$k$-deg(add)}~\cite{liu2008towards}. This anonymization method ensures that $k$-degree anonymity is preserved by only adding edges. 
	\item \textbf{$k$-degree(add \& del)}~\cite{liu2008towards}. This method ensures that $k$-degree anonymity is preserved by performing simultaneous add/removal of the edges.
	
	\item \textbf{Switching}. This method selects two random edges $\left(i_1, j_1\right)$ and
	$\left(i_2, j_2\right)$ from the original graph such that $\{\left(i_1, j_2\right) \notin \mathcal{E} \wedge \left(i_2, j_1\right)\notin \mathcal{E}\}$. Then, it switches pairs of edges, i.e. remove edges $\left(i_1, j_1\right)$ and
	$\left(i_2, j_2\right)$ and add new edges $\left(i_1, j_2\right)$ and $\left(i_2, j_1\right)$ instead. This step is repeated $\frac{p|\mathcal{E}|}{2}$ times which results in $p|\mathcal{E}|$ edge removals/additions.
	
	\item \textbf{Perturbation}. This method is also known as \textit{unintended} anonymziation and has two main steps. It first removes $p|\mathcal{E}|$ edges in a same way as sparsification method does. Then, it adds random false edges until the number of edges in the anonymized graph is the same as the original one.
\end{itemize}

Furthermore, the \textit{Textual information} is anonymized using the techniques discussed earlier in Section \ref{text-anonymiziation} as follows:

\begin{itemize}[leftmargin=*]
	\item \textbf{Naive Anonymization}. This approach first removes users' identifiers and links from the tweets and then vectorize it.
	\item \textbf{Diff Privacy}. This method takes the output of the naive anonymization technique and then ensures differential privacy by adding Laplacian noise to the generated text vector.
\end{itemize}
\vspace{-10pt}
\subsection{Results and Discussion}
\subsubsection{Experimental Settings}
We evaluate de-anonymization approaches by a metric called \textit{success rate} $\mathcal{X}=\frac{n_c}{N}$, where $n_c$ is the total number of users that have been successfully re-identified and $N$ is the total number of users in the anonymized dataset!\cite{narayanan2009anonymizing}. Larger values of this measure correspond to higher privacy breach.

Following the previous works~\cite{fu2015effective,qian2016anonymizing}, we set $k=10$ for $k$-degree anonymity and $p=0.1$ for sparsification, purturbation and switching methods. The $\epsilon$ for differential privacy technique is set as $\epsilon=0.01$. We also set the parameters of \textsc{Athd} as follows: $\{k=10, \alpha=0.5, \beta=0.7, \lambda=20, \theta=50, b = 7, \delta = 50\}$. The values of $\delta$ and $b$ for quantifying degree distributions are chosen such that it can accommodate higher degrees variation. Empirical results showed that the choice of $\delta$ and $b$ does not have a huge impact on the final results. We also set the number of returned profiles as $h=1$. Clearly, increasing the value of $h$ will increase the de-anonymization success rate. To answer the research questions, we make 12 copies of the original data and sanitize each copy with a different combination of structural and textual anonymization techniques discussed earlier. For evaluation, we define two different variants of our proposed approach, \textsc{Athd}, as follows:
\begin{itemize}[leftmargin=*]
	\item \textbf{\textsc{Athd}-Simple:} This uses Eq.\ref{simple1} and Eq.\ref{simple2} to calculate similarity.
	\item \textbf{\textsc{Athd}-Improved:} This variant uses Eq.\ref{TotalSim} to improve similarity measure by incorporating features from neighbors to reduce the anonymization effect.
\end{itemize}
\vspace{-10pt}
\subsubsection{Perfomrnace Comparison}
To evaluate the effectiveness of \textsc{Athd}, we benchmark its two variants, \textsc{Athd}-Simple and \textsc{Athd}-Improved, against the following two baselines. 
\begin{itemize}[leftmargin=*]
	\item \textbf{Narayanan et. al.}~\cite{narayanan2009anonymizing}: It computes the similarity between an unmapped user $u$ and a candidate $c_i$, by using the number of neighbors of $u$ that have been mapped to neighbors of $c_i$.
	\item \textbf{ADA}~\cite{ji2016general}: This method considers a combination of structural, relative distance and inheritance similarity. We only use degree centrality for measuring structural similarity as we do not have access to the global structure of $c_i$ in $\mathcal{T}$.
\end{itemize}

In general, these baselines are seed-based approaches, meaning that they map a known target user $v$ in $\mathcal{T}$ to a user in the anonymized data by utilizing a small set of initially mapped seed users and then propagating the mappings through the whole data. These works also need a previously collected background knowledge $\mathcal{B}$. We need to use same settings to make a fair comparison between the baselines and our proposed framework. To do so, we first make an initial seed set of the size $\nu=20$, by mapping a set of random users in the anonymized dataset to their real identities for each of Twitter and Foursquare data. Then, we repeat the same 3-step procedure as in the \textsc{Athd} for the baselines, except that we the similarity metric in the last step is replaced with those of the baselines.  
 Performance comparison results for both datasets are demonstrated in Table \ref{Res2} with the following observations:


\begin{itemize}[leftmargin=*]
	\item Narayanan et. al. is the least effective de-anonymization on both datasets. The reason is because its utilized similarity metric relies on the set of previously mapped neighbors and ignores the available structural and textual information provided in the data.
	\item ADA approach is more powerful than Narayanan et. al. since it incorporates structural properties of the data.
	\item Anonymized data is more vulnerable to \textsc{Athd}-Simple compared to ADA and Narayanan et. al. This is because both structural and textual information are incorporated in the similarity metric used in \textsc{Athd}-Simple. This confirms that integrating different components of data plays an important role in de-anonymization for heterogeneous social media data. 
	\item \textsc{Athd}-Improved technique achieves the best results for both Twitter and Foursquare datasets. This demonstrates the effectiveness of utilizing homophily and the features of neighbors for more effective de-anonymization.
\end{itemize}

To recap, the above observations confirm the efficiency of our proposed approach \textsc{Athd}.
\vspace{-0.25cm}
\subsubsection{Assessing Effectiveness of Anonymization}
Having discussed the efficiency of the proposed \textsc{Athd} de-anonymization approach, we now seek the answer to the last two questions. The performance results w.r.t. the four anonymization cases are demonstrated in Table \ref{Res2}. The numbers in parentheses demonstrate the corresponding case number defined earlier in the introduction. We make the following observations for both datasets:

\begin{itemize}[leftmargin=*]
	\item Publishing the data with no anonymization for either aspect (i.e., case 1) resulted in a large information breach in both \textsc{Athd}-Simple and \textsc{Athd}-Improved approaches which suggests the least amount of protection as expected.
	\item In general, anonymizing either aspect of the data (i.e., cases 2 and 3) protects users privacy more than case 1.
	\item Case 4 is the strongest protection among the four cases. Accordingly, the answer to the second question is no.
	\item  Although case 4 provides the strongest protection, \textsc{Athd}-Improved was able to re-identify at least $56\%$ of the users in the anonymized dataset, which is a significant number in the field of privacy. This shows that case 4 is far from sufficient for data anonymization. 
	\item Sparsification is the most vulnerable anonymization approach against both \textsc{Athd}-Simple and \textsc{Athd}-Improved techniques as it makes the least amount of changes to the link information.
	\item Although the switching and perturbation methods both add and deletes the same number of edges, switching is more vulnerable to the de-anonymization since it preserves the node degrees.
	\item Despite the fact that $k$-degree anonymity based approaches guarantee the user re-identification probability to be at most $\frac{1}{k}$, but they fail because of using extra textual information.
\end{itemize}

According to these observations, the answers to the introduced research questions are no. These results further indicate that despite anonymization of all aspects of data is essential, but it is not sufficient to anonymize each aspect independently from others. This is because an adversary could easily breach privacy no matter what anonymization algorithm has been used. Consequently, serious privacy breach could happen when the published data is heterogeneous. This necessitates taking into account the latent relations in different portions of the social media data for anonymization.
\vspace{-10pt}
\section{Related Work}\label{related}


\textbf{Social Network Anonymization}. Social networks contain private profile information and sensitive social relationships which provide opportunities for researchers to study and understand individuals at unprecedented scales~\cite{beigi2016signed,hajibagheri2017learning,hajibagherichapter,gbeigi}. However, this information may leak users' privacy~\cite{backstrom2007wherefore}. Anonymization methods serve as an important role to maintain data utility as well as protecting privacy~\cite{wu2010survey}. Existing social network anonymization methods can be categorized mainly into three categories: \textit{$k$-anonymity}, \textit{edge randomization}, \textit{clustering-based generalization} and \textit{differential privacy}. The aim of $k$-anonymity methods is to anonymize each node so that it is indistinguishable from at least $k-1$ other nodes~\cite{sweeney2002k}. Liu~\textit{et al.} proposed to achieve $k$-degree anonymization~\cite{liu2008towards} through edge addition/deletion strategies~\cite{liu2008towards}. 
Zhou~\textit{et al.} further considered the assumption that the adversary knows subgraph constructed by the immediate neighbors of a target node, and aims to achieve $k$-neighborhood anonymity~\cite{zhou2008preserving}. 
Edge randomization algorithms for social networks usually utilize edge-based randomization strategies to anonymize data, such as random adding/deleting and random switching~\cite{ying2009graph}. Clustering-based anonymization methods group nodes and edges, and only reveal the density and size so that individual attributes are protected~\cite{tassa2013anonymization}. Another work seeks to generate an anonymized graph which guarantees differential privacy~\cite{sala2011sharing}.
\textbf{Social Network De-anonymization.} De-anonymization approaches on social networks aim to re-identify the anonymous user data by using previously collected background information. Existing de-anonymization methods can be categorized into i) \textit{seed-based} and ii) \textit{seed-free}, according to whether pre-annotated seed users exist or not. Seed-based de-anonymization attack on social network was proposed to use only structural information and propagates node mappings based on seed user pairs~\cite{narayanan2009anonymizing}.  Later, Narayanan~\textit{et al.} ~\cite{narayanan2011link} employed a simplified attack using less heuristics rules for link prediction problem. Nilizadeh~\textit{et al.} further proposed a community-enhanced de-anonymization scheme, which first de-anonymizes data in community-level and then de-anonymizes the users within the communities~\cite{nilizadeh2014community}. Yartseva~\textit{et al.} proposed a percolation-based de-anonymization method using neighborhood overlap information~\cite{yartseva2013performance}. Seed-free approaches assume there is no seed users available. Pedarsani~\textit{et al.} presented a Bayesian model to iteratively perform a maximum weighted bipartite graph matching starting from the nodes with the highest degree~\cite{pedarsani2013bayesian}. Moreover, Ji~\textit{et al.} proposed to use optimization based methods to minimize the edge difference between anonymized network and background information~\cite{ji2014structural}. Recently, another group of works have focused on exploiting additional sources of information such as profile information~\cite{fu2015effective} and users attributes~\cite{qian2016anonymizing} for social graph de-anonymization. Fu~\textit{et al.} proposed to use structural and descriptive information to de-anonymize users without seed nodes~\cite{fu2015effective}. A thorough survey on graph data anonymization and de-anonymization is presented in~\cite{ji2016graph}. Note that de-anonymization methods are similar to those of user identity linkage across social network when only network information is available~\cite{shu2017user}. In addition to the different goals of these two research direction, the main difference is that the given graph structured is not anonymized in case of user identity linkage problem. This makes the de-anonymization much more challenging.
There are two main differences between our work and the above works. First, we introduce a new adversarial attack specialized for social media data and second, we assess the efficiency of existing anonymization techniques for heterogeneous social media data.
\vspace{-10pt}
\section{Conclusion}\label{concl}
In this work, we study a new problem of user data privacy for social media via an adversarial approach. Our work differs from the existing works due to unique properties of social media data: a social media site has an inordinate number of users and the site only allows for a limited number of data queries. Since anonymization takes time and requires dedicated efforts, anonymization efficiency should be maximized. Thus, we evaluate the strengths of anonymization techniques in the context of social media data and verify if it is sufficient. We propose \textsc{Athd}, a novel adversarial technique by exploiting heterogeneous characteristics of social media data. Our results illustrate that anonymizing even all aspects of data is not sufficient for protecting user privacy due to hidden relations between different aspects of the heterogeneous data. 
One future research direction for our work is to examine how different combinations of heterogeneous data (e.g., a combination of location and textual information) are vulnerable to the de-anonymization attack, though the work reported in this paper is sufficient to show the need for better anonymization with resource constraints.  Another potential direction is to improve anonymization techniques to preserve the privacy of users in social media data by considering hidden relations between different components of the data due to the innate heterogeneity of user-generated data. 
\vspace{-10pt}
\begin{acks}
	The authors would like to thank Alexander Nou for his help throughout the paper. This material is based upon the work supported in part by Army Research Office (ARO) under grant number W911NF-15-1-0328 and Office of Naval Research (ONR) under grant number N00014-17-1-2605.
\end{acks}


\end{document}